\documentclass{PoS}
\usepackage{amsmath,amssymb,bm,longtable,mathrsfs,slashed}

\title{\begin{picture}(0,0)(0,0)%
   \put(360,68){\makebox(0,0)[l]{\textnormal{\normalsize OU-HET-1023}}}%
\end{picture}Axial U(1) symmetry, topology, and Dirac spectra at high temperature in $N_f=2$ lattice QCD}

\ShortTitle{Axial U(1) symmetry, topology, and Dirac spectra at high temperature in $N_f=2$ lattice QCD}

\author{\speaker{Kei Suzuki}$\, ^a$\thanks{E-mail: kei.suzuki@kek.jp} , Sinya Aoki$\, ^b$, Yasumichi Aoki$\, ^{c}$\thanks{current address: KEK$^a$ and RIKEN Center for Computational Science, Kobe 650-0047, Japan.}, Guido Cossu$\, ^d$, Hidenori Fukaya$\, ^e$, Shoji Hashimoto$\, ^{a,f}$ \ (JLQCD Collaboration)\\
        \llap{$^a$}KEK Theory Center, High Energy Accelerator Research Organization (KEK), Tsukuba 305-0801, Japan\\
        \llap{$^b$}Center for Gravitational Physics, Yukawa Institute for Theoretical Physics, Kyoto 606-8502, Japan\\
        \llap{$^c$}RIKEN BNL Research Center, Brookhaven National Laboratory, Upton, NY 11973, USA\\
        \llap{$^d$}School of Physics and Astronomy, The University of Edinburgh, Edinburgh EH9 3JZ, United Kingdom\\
        \llap{$^e$}Department of Physics, Osaka University, Toyonaka 560-0043, Japan\\
        \llap{$^f$}School of High Energy Accelerator Science, The Graduate University for Advanced Studies (Sokendai), Tsukuba 305-0801, Japan\\
}

\abstract{%
Using lattice QCD simulations with $N_f = 2$ dynamical fermions, we study the axial $U(1)$ symmetry, topological charge, and Dirac eigenvalue spectra in the high-temperature phase in which the chiral symmetry is restored.
Our gauge ensembles are generated with M\"obius domain-wall fermions, but the measurements such as susceptibilities are reweighted to those for the overlap fermions by using overlap/domain-wall reweighting technique.
We find that the $U(1)_A$ and topological susceptibilities are strongly suppressed in the small quark mass region, which is related to the reduction of chiral-zero and low-nonzero modes on the Dirac spectra. 
We also examine their volume dependence.
}

\FullConference{The 9th International workshop on Chiral Dynamics\\
		17-21 September, 2018\\
		Durham, NC, USA.}

\begin{document}
\section{Introduction}\label{sec-1}
The axial $U(1)_A$ symmetry plays a uniquely important role in quantum chromodynamics (QCD).
In the low-temperature phase of QCD, it is violated by the chiral anomaly.
This anomaly is closely related to topological excitations of background gluon fields, such as the instantons, and it induces the larger mass of the $\eta^\prime$ meson.
In the high-temperature phase, the (spontaneously broken) chiral symmetry is restored while the fate of the $U(1)_A$ symmetry is a longstanding problem of QCD and is still under debate.

\begin{figure}[tb!]
    \begin{minipage}[t]{1.0\columnwidth}
        \begin{center}
            \includegraphics[clip, width=1.0\columnwidth]{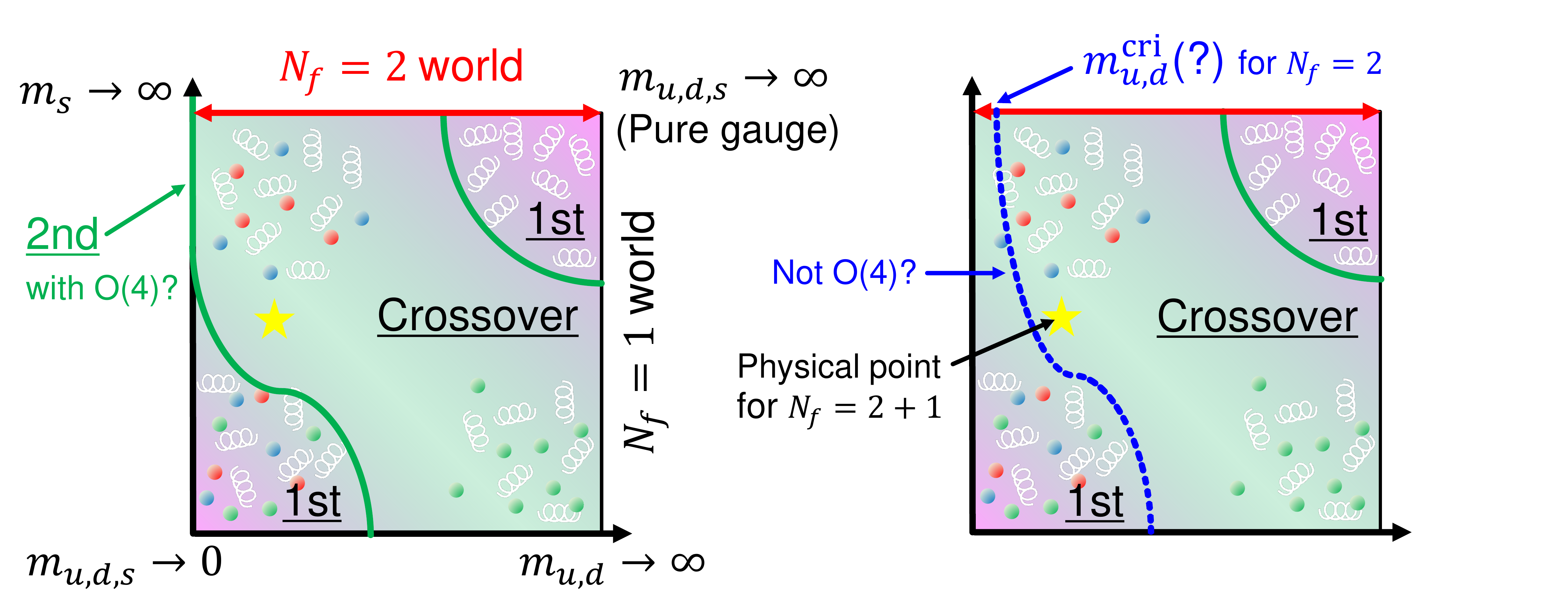}
        \end{center}
    \end{minipage}
    \caption{Phase diagrams of QCD for up and down quark masses $m_{u,d}$ (horizontal axis) and strange quark mass $m_s$ (vertical axis).
Left: The conventional phase diagram.
Right: A possible diagram when the $U(1)_A$ is restored above $T_c$, which is suggested in Ref.~\cite{Aoki:2012yj}.}
    \label{Fig-col}
\end{figure}

This problem has been attacked using lattice QCD simulations at $N_f=2$ \cite{Cossu:2013uua,Chiu:2013wwa,Brandt:2016daq,Tomiya:2016jwr} and $N_f=2+1$ \cite{Bazavov:2012qja,Buchoff:2013nra,Bhattacharya:2014ara,Dick:2015twa,Mazur:2018pjw} as well as analytic approaches.
In particular, in Ref.~\cite{Cohen:1996ng}, Cohen suggested that the $U(1)_A$ symmetry of massless $N_f=2$ QCD is restored if the contributions from the zero modes of Dirac eigenvalues can be ignored.
As a result, the mesonic correlators for $\pi$, $\sigma$, $\delta$, and $\eta$ channels can degenerate.
More recently, the authors of Ref.~\cite{Aoki:2012yj} analytically proved that the violation of the $U(1)_A$ symmetry of massless $N_f=2$ QCD becomes invisible under some assumptions such as the analyticity of the Dirac spectral density near the origin (for an alternative proof, see Ref.~\cite{Kanazawa:2015xna}).
They also suggested a possible modification of the phase diagram for up and down quark masses $m_{u,d}$ and strange quark mass $m_s$ (which is known as the so-called Columbia plot, see Fig.~\ref{Fig-col}), based on an effective theory \cite{Pisarski:1983ms}.
When the $U(1)_A$ symmetry is violated in the chiral limit ($m_{u,d}=0$) for $N_f=2$, the chiral phase transition at $m_{u,d}=0$ is expected to be second-order belonging to the three-dimensional $O(4)$ universality class (see the left of Fig.~\ref{Fig-col}).
On the other hand, if the $U(1)_A$ symmetry is restored, the chiral phase transition at $m_{u,d}=0$ may be first-order or second-order belonging to an unusual universality class (which is not $O(4)$).
In the case of first-order transition, we expect that a nonzero ``critical mass'' $m_{u,d}^\mathrm{cri}$ may exist since the first-order region ($m<m_{u,d}^\mathrm{cri}$) and the crossover region ($m>m_{u,d}^\mathrm{cri}$) must be separated by a mass boundary (see the right of Fig.~\ref{Fig-col}).
The existence of such a critical mass could also affect even the phase structure of $N_f=3$ QCD.
In addition, for other theoretical discussions, see Refs.~\cite{Pelissetto:2013hqa,Nakayama:2014sba,Sato:2014axa,Nakayama:2016jhq,Azcoiti:2016zbi,Azcoiti:2017jsh,GomezNicola:2017bhm,Nicola:2018vug}.

Employing chiral symmetric fermion actions, the JLQCD Collaboration observed a restoration of the $U(1)_A$ symmetry above the critical temperature $T_c$ in $N_f=2$ lattice QCD~\cite{Cossu:2013uua,Tomiya:2016jwr}.
In Ref.~\cite{Cossu:2013uua}, the $U(1)_A$ symmetry was investigated using the Dirac eigenvalue spectra on gauge configurations generated with the dynamical OV fermions in a fixed topological sector, $Q=0$.
After that, in Ref.~\cite{Tomiya:2016jwr}, the gauge configurations with dynamical M\"obius domain-wall (MDW) fermions \cite{Brower:2005qw,Brower:2012vk} has bee used without fixing topological sectors.
Since the GW relation for MDW fermions is slightly violated especially for larger lattice spacings \cite{Cossu:2015kfa}, we utilized the DW/OV reweighting technique \cite{Fukaya:2013vka,Tomiya:2016jwr}.
In this case, an observable measured on the gauge ensembles with dynamical MDW fermions can be transformed (or reweighted) to that on OV fermion ensembles, and then we can finally evaluate the observable satisfying the GW relation.

In these proceedings, in order to examine the $U(1)_A$ symmetry in the high-temperature phase above $T_c$, we show the results of the $U(1)_A$ susceptibility, topological susceptibility, and corresponding Dirac spectra from $N_f=2$ lattice QCD simulations, where finer lattice spacing, $1/a=2.64 \, \mathrm{GeV}$ ($a \sim 0.075 \, \mathrm{fm}$), than the previous works~\cite{Cossu:2013uua,Tomiya:2016jwr} is used.
Note that a part of the results has also been reported in previous proceedings~\cite{Aoki:2017xux,Suzuki:2017ifu,Suzuki:2018vbe}.

\section{Simulation setup}\label{sec-2}

\subsection{$U(1)_A$ susceptibility on the lattice}\label{subsec-2-1}
As an order parameter of the $U(1)_A$ symmetry breaking, the $U(1)_A$ susceptibility, $\Delta_{\pi - \delta}$, is defined as a difference between the correlators of isovector-pseudoscalar ($\pi^a \equiv i \bar{\psi} \tau^a \gamma_5 \psi$) and isovector-scalar ($\delta^a \equiv \bar{\psi} \tau^a \psi$) operators: 
\begin{equation}
\Delta_{\pi-\delta} \equiv \chi_\pi - \chi_\delta \equiv \int d^4x \langle \pi^a(x) \pi^a(0) - \delta^a (x) \delta^a(0) \rangle, \label{eq:Delta_def}
\end{equation}
where $a$ is an isospin index when we consider two-flavor ($N_f=2$) QCD.

In the continuum theory, the $U(1)_A$ susceptibility (\ref{eq:Delta_def}) can be rewritten by the Dirac eigenvalue spectral density for fermions with a mass $m$:
\begin{equation}
\Delta_{\pi-\delta} = \int_0^\infty d\lambda\,\rho(\lambda) \frac{2m^2}{(\lambda^2+m^2)^2}, \label{eq:Delta_cont}
\end{equation}
where $\rho(\lambda)=(1/V)\langle\sum_{\lambda'}\delta(\lambda-\lambda')\rangle$ with the Dirac eigenvalues $\lambda$ is the Dirac eigenvalue spectral density, and $V=L^3\times L_t$ is the four-dimensional volume.
In the lattice theory, the $U(1)_A$ susceptibility for OV fermion operators is given by \cite{Cossu:2015kfa}
\begin{equation}
\Delta_{\pi-\delta}^{\mathrm{ov}} =  \frac{1}{V(1-m^2)^2} \left< \sum_i \frac{2m^2(1-\lambda_i^{(\mathrm{ov},m)2})^2}{\lambda_i^{(\mathrm{ov},m)4}} \right> , \label{eq:Delta_ov}
\end{equation}
where $\lambda_i^{(\mathrm{ov},m)}$ is the $i$-th eigenvalue of the (hermitian) massive overlap-Dirac operator, and the lattice spacing is set to $a=1$.
If the GW relation is not exact, we have to introduce additional terms in Eq.~(\ref{eq:Delta_ov}) \cite{Cossu:2015kfa}.

In the following, we discuss two types of subtractions in the $U(1)_A$ susceptibility: the chiral zero modes and the ultraviolet divergence should be subtracted by a definition. 

The definition of Eq.~(\ref{eq:Delta_ov}) contains the contribution from nontrivial topological sectors, which is related to chiral zero modes with $\lambda_i^{(\mathrm{ov},m)} \approx \pm m$, where ``$\approx$'' means possible small violation of the GW relation in our setup (If the GW relation is exact, then $\lambda_i^{(\mathrm{ov},m)} = \pm m$ holds).
After subtracting zero modes, we define a modified $U(1)_A$ susceptibility:
\begin{equation}
\bar{\Delta}_{\pi-\delta}^{\mathrm{ov}} \equiv \Delta_{\pi-\delta}^{\mathrm{ov}} - \frac{1}{V(1-m^2)^2} \left< \sum_{0-mode} \frac{2m^2(1-\lambda_i^{(\mathrm{ov},m)2})^2}{\lambda_i^{(\mathrm{ov},m)4}} \right> . \label{eq:Delta_bar}
\end{equation}
In the thermodynamic limit ($V \to \infty$), such a subtraction of zero modes can be justified as follows \cite{Aoki:2012yj}.
If the GW relation is exact, the second term of Eq.~(\ref{eq:Delta_bar}) can be written as $2N_0/Vm^2$, where $N_0$ is the number of chiral zero modes.
$\langle N_0^2 \rangle$ is expected to scale as $O(V)$, so that $N_0/V$ also scale as $O(1/\sqrt{V})$.
Therefore, the contribution from the chiral zero modes vanishes in the thermodynamic limit: $N_0/V \to 0$ as $V \to \infty$.

Next, we subtract contributions from the ultraviolet divergence.
Eq.~(\ref{eq:Delta_cont}) in the continuum theory contains an (logarithmic) ultraviolet divergence.
Eq.~(\ref{eq:Delta_ov}) on the lattice includes a large contribution from the lattice cutoff $\Lambda$ instead of the divergence, which is proportional to $m^2 \ln \Lambda$.
Therefore, $\Delta_{\pi-\delta}(m)$ at a valence quark mass $m$ can be parametrized as
\begin{equation}
\Delta_{\pi-\delta} (m) = \frac{a}{m^2} + b + c m^2 + \mathcal{O} (m^4), \label{eq:Delta_parametrized}
\end{equation}
where the first term is the contribution from the chiral zero modes.\footnote{When the GW relation is exact, the second term of Eq.~(\ref{eq:Delta_bar}) is equivalent to the $\frac{a}{m^2}$ term of Eq.~(\ref{eq:Delta_parametrized}).
In that sense, the procedure of Eqs.~(\ref{eq:Delta_parametrized}) and (\ref{eq:Delta_subt}) is an alternative method to subtract zero modes.}
The second term is the $U(1)_A$ violation we desire.
The third term represents the contribution of $m^2 \ln \Lambda$.
Here, in order to remove $a$ and $c$, and to get only $b$, we use three susceptibilities $\Delta_{\pi-\delta} (m_{1,2,3})$ for three different valence quark masses ($m_2<m_1<m_3$, where for example we choose $m_2=0.999 m_1$ and $m_3=1.001 m_1$ to keep the partially quenching artifact under control):
\begin{equation}
b \simeq \Delta_{\pi-\delta}^\mathrm{finite} \equiv \frac{(m_1^2+m_2^2)(m_1^2+m_3^2)}{m_3^2-m_2^2} \left[\frac{m_1^2 \Delta(m_1) -m_2^2 \Delta(m_2)}{m_1^4 - m_2^4} -\frac{m_1^2 \Delta(m_1) - m_3^2 \Delta(m_3)}{m_1^4-m_3^4} \right]. \label{eq:Delta_subt}
\end{equation}
Note that we can use this formula also for $\bar{\Delta}_{\pi-\delta} (m)$ as defined in Eq.~(\ref{eq:Delta_bar}).
In that case, the contributions from the zero modes in $\Delta_{\pi-\delta}^\mathrm{finite}$ are subtracted by Eq.~(\ref{eq:Delta_bar}), so that the contribution with $a/m^2$ is already absent.

\subsection{Topological susceptibility on the lattice}\label{subsec-2-2}
The topological susceptibility is defined as the gauge ensemble average
\begin{equation}
\chi_t=\frac{\langle Q_t^2 \rangle}{V}, \label{chit}
\end{equation}
where the topological charge $Q_t$ is an integer value and we have two types of definitions.
As a fermionic definition, it is determined by the index theorem for the overlap Dirac operator:
\begin{equation}
Q_t=n_+ - n_-, \label{Q_fermion}
\end{equation}
where $n_\pm$ is the number of chiral zero modes with positive or negative chirality.
As a gluonic definition, the topological charge at flow time $t$ is defined as the summation over the spacetime $x$:
\begin{equation}
Q_t (t)=\frac{1}{32\pi^2} \sum_x \varepsilon^{\mu \nu \rho \sigma} \mathrm{Tr} \, F_{\mu \nu}(x,t) F_{\rho \sigma} (x,t), \label{Q_gluon}
\end{equation}
where $F_{\mu \nu}$ is the field strength defined by clover construction on the lattice \cite{Bruno:2014ova}.
This definition is generally not an integer, but we see a clear well-discretized distribution of $Q_t$ at $t=5$. We round off its non-interger part.

\subsection{Numerical setup}\label{subsec-2-3}
We use the lattice with the imaginary time length $L_t =12$ which corresponds to $T=220 \, \mathrm{MeV}$ at the lattice spacing, $1/a=2.64 \, \mathrm{GeV}$ ($a \sim 0.075 \, \mathrm{fm}$).
To carefully examine finite volume effects, we apply the spatial lengths $L=24,32,40,48$. 
Also, to study quark mass dependence, we use the five kinds of quark masses: $am=0.001-0.01$ ($2.64-26.4 \, \mathrm{MeV}$).\footnote{Here, $m$ is the bare quark mass and is not renormalized.
We estimate a preliminary value of the physical quark mass (the average of the up and down quarks) to be $am = 0.0014(2)$ ($3.7(5) \, \mathrm{MeV}$).}
In Table \ref{Tab:param}, we summarize the simulation parameters.

\begin{table}[t!]
  \small
  \centering
\caption{Numerical parameters in lattice simulations.
$L^3 \times L_t $, $L_s$, $\beta$, $a$, and $m$ are the lattice size, length of the fifth dimension in the M\"obius domain-wall fermion,
gauge coupling, lattice spacing, and quark mass, respectively.
}
\begin{tabular}{cccccc}
\hline\hline
$L^3 \times L_t $ & $L_s$ & $\beta$ & $a$ [fm] & $T$ [MeV] & $am$ \\
\hline
$24^3 \times 12$ & 16 & 4.30 & 0.075 & 220 & 0.001   \\
$24^3 \times 12$ & 16 & 4.30 & 0.075 & 220 & 0.0025  \\
$24^3 \times 12$ & 16 & 4.30 & 0.075 & 220 & 0.00375 \\
$24^3 \times 12$ & 16 & 4.30 & 0.075 & 220 & 0.005   \\
$24^3 \times 12$ & 16 & 4.30 & 0.075 & 220 & 0.01    \\
\hline
$32^3 \times 12$ & 16 & 4.30 & 0.075 & 220 & 0.001   \\
$32^3 \times 12$ & 16 & 4.30 & 0.075 & 220 & 0.0025  \\
$32^3 \times 12$ & 16 & 4.30 & 0.075 & 220 & 0.00375 \\
$32^3 \times 12$ & 16 & 4.30 & 0.075 & 220 & 0.005   \\
$32^3 \times 12$ & 16 & 4.30 & 0.075 & 220 & 0.01    \\
\hline
$40^3 \times 12$ & 16 & 4.30 & 0.075 & 220 & 0.005   \\
$40^3 \times 12$ & 16 & 4.30 & 0.075 & 220 & 0.01    \\
\hline
$48^3 \times 12$ & 16 & 4.30 & 0.075 & 220 & 0.001   \\
$48^3 \times 12$ & 16 & 4.30 & 0.075 & 220 & 0.0025  \\
$48^3 \times 12$ & 16 & 4.30 & 0.075 & 220 & 0.00375 \\
$48^3 \times 12$ & 16 & 4.30 & 0.075 & 220 & 0.005   \\
\hline\hline
\end{tabular}
\label{Tab:param}
\end{table}

We use the tree-level Symanzik improved gauge action.
For the fermion part, we apply the MDW fermions \cite{Brower:2005qw,Brower:2012vk} with a smeared link.
By using the DW/OV reweighting technique \cite{Fukaya:2013vka,Tomiya:2016jwr}, an observable $\mathcal{O}$ measured on the MDW fermion ensembles is transformed to that on the OV fermion:
\begin{equation}
\langle \mathcal{O} \rangle_{\mathrm{ov}} = \frac{ \langle \mathcal{O} R\rangle_{\mathrm{DW}} }{ \langle R\rangle_{\mathrm{DW}}},
\end{equation}
where the two types of expectation values, $\langle \cdots \rangle_{\mathrm{DW}}$ and $\langle \cdots \rangle_{\mathrm{ov}}$, are the ensemble average with the MDW and reweighted OV fermions, respectively.
A value $R$ is called the reweighting factor and it is stochastically estimated on the MDW fermion ensembles \cite{Fukaya:2013vka,Tomiya:2016jwr}.
The reweighting procedure reduces the violation of the GW relation for the MDW fermions.
\section{Preliminary results}\label{sec-3}

\begin{figure}[t!]
    \centering
    \begin{minipage}[t]{1.0\columnwidth}  
            \includegraphics[clip,width=1.0\columnwidth]{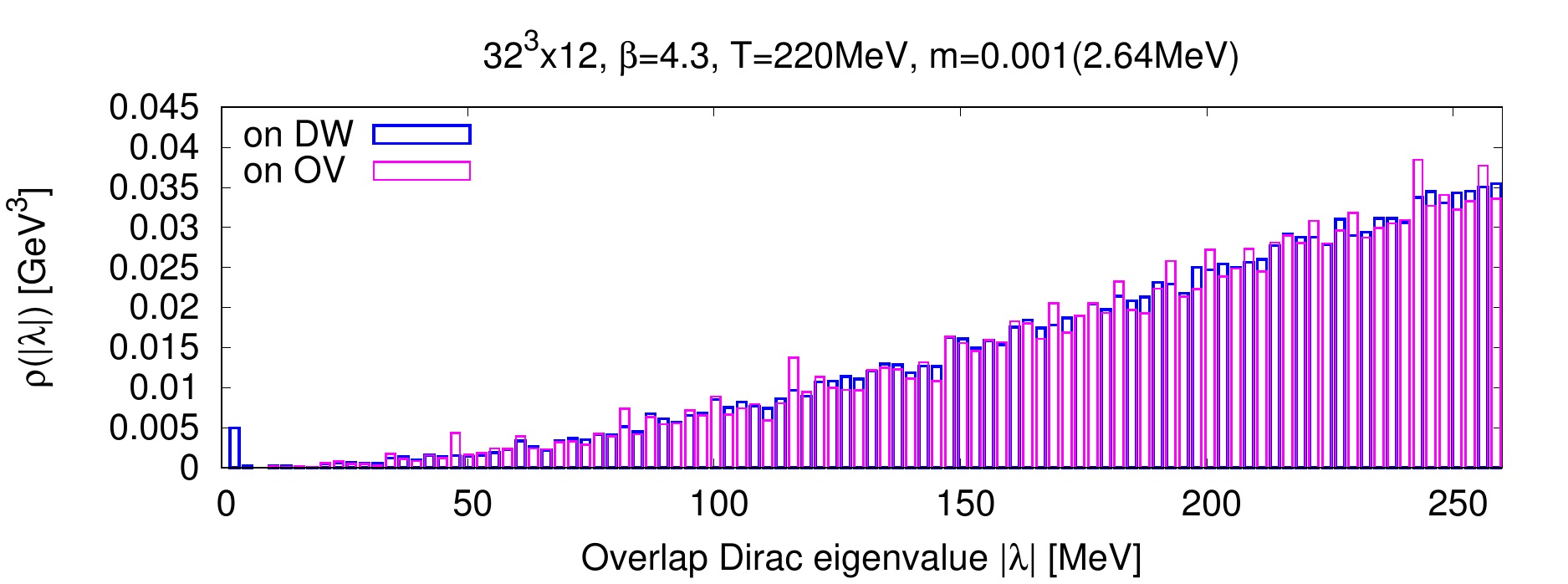}       
    \end{minipage}
    \begin{minipage}[t]{1.0\columnwidth}
            \includegraphics[clip,width=1.0\columnwidth]{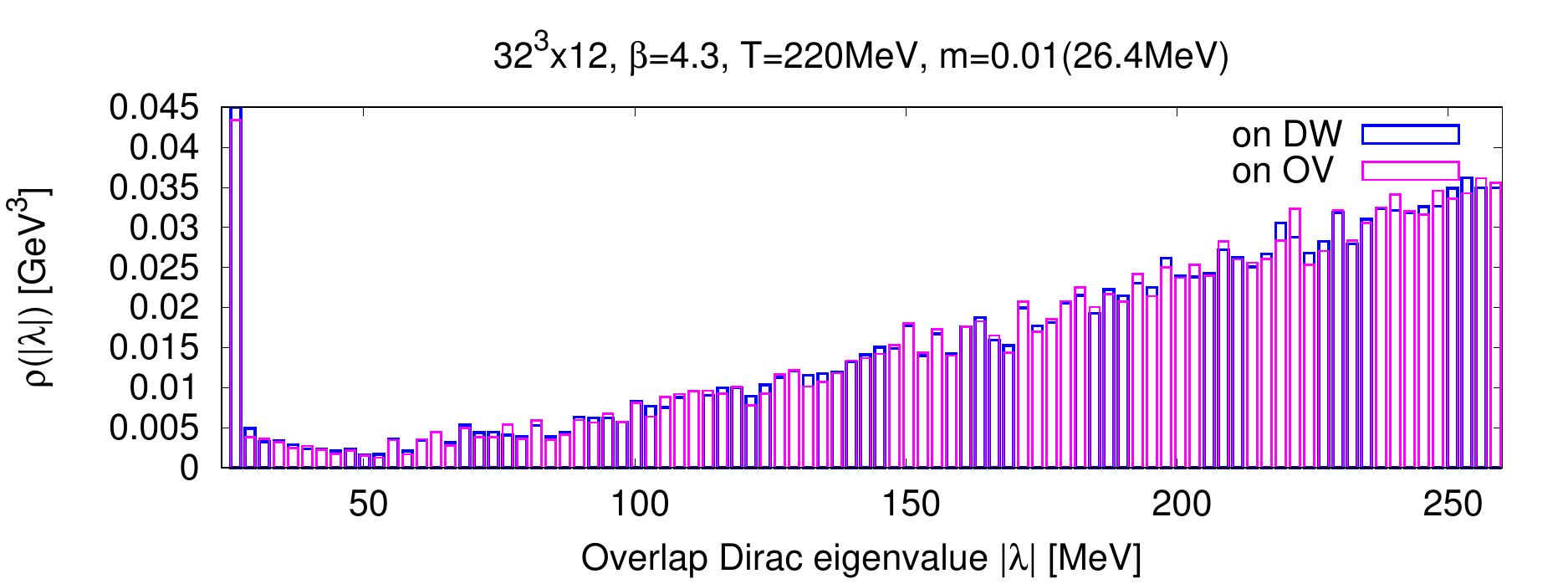}
    \end{minipage}
    \caption{Spectral density $\rho(|\lambda|)$ for overlap-Dirac eigenvalues $\lambda$ at $T=220 \, \mathrm{MeV}$.
Upper panel: $m=2.64 \, \mathrm{MeV}$.
Lower panel: $m=26.4 \, \mathrm{MeV}$.
Blue and magenta bins correspond to the spectra on the original M\"obius domain-wall (DW) and reweighted overlap (OV) fermion ensembles, respectively.
 }
    \label{fig-1}
\end{figure}

\subsection{Spectral density of overlap-Dirac eigenvalues}\label{subsec-3-1}
In Fig.~\ref{fig-1}, we show the spectral density $\rho(|\lambda|)$ of the overlap-Dirac eigenvalues $\lambda$ at $T=220 \, \mathrm{MeV}$, which is observed both on the MDW ensembles (blue bins) and reweighted OV (magenta bins) ensembles.
As shown in the upper panel of Fig.~\ref{fig-1}, at a light quark mass $m=2.64 \, \mathrm{MeV}$, we find that the eigenmodes in the low energy region are strongly suppressed.
Then the chiral zero modes and higher nonzero modes can be clearly distinguished.
According to the definition (\ref{eq:Delta_bar}) of the $U(1)_A$ susceptibility, $\bar{\Delta}_{\pi-\delta}^{\mathrm{ov}}$ is induced by the {\it low} non-zero modes (chiral zero modes are subtracted by definition).
Therefore, the strong suppression of the low modes on the spectra leads to the small value of $\bar{\Delta}_{\pi-\delta}^{\mathrm{ov}}$.
Notice that the zero modes on the spectrum on the DW (blue bins) are caused by the discrepancy between the valence (OV) quark and sea (MDW) quarks.
In other words, these zero modes are artifacts induced by the partially quenched approximation.
After the DW/OV reweighting (magenta bins), we can completely remove such fictitious zero modes.

As shown in the lower panel of Fig.~\ref{fig-1}, at a large quark mass $m=26.4 \, \mathrm{MeV}$, not only the chiral zero modes but also low nonzero modes are enhanced more frequently.
Then we cannot clearly separate the zero modes from other modes.
The increase of the low but nonzero modes leads to a large value of $\bar{\Delta}_{\pi-\delta}^{\mathrm{ov}}$, as shown in Subsection \ref{subsec-3-2}.
Here, the zero modes observed on the DW ensemble survive even after the DW/OV reweighting, which indicates that these zero modes are not artifacts but really physical ones.
The appearance of these physical zero modes is related to the nonzero values of the topological charge and susceptibility, as discussed in Subsection \ref{subsec-3-3}.

\subsection{$U(1)_A$ susceptibility}\label{subsec-3-2}
\begin{figure}[t!]
    \begin{minipage}[t]{0.5\columnwidth}
    \begin{center}
            \includegraphics[clip,width=1.0\columnwidth]{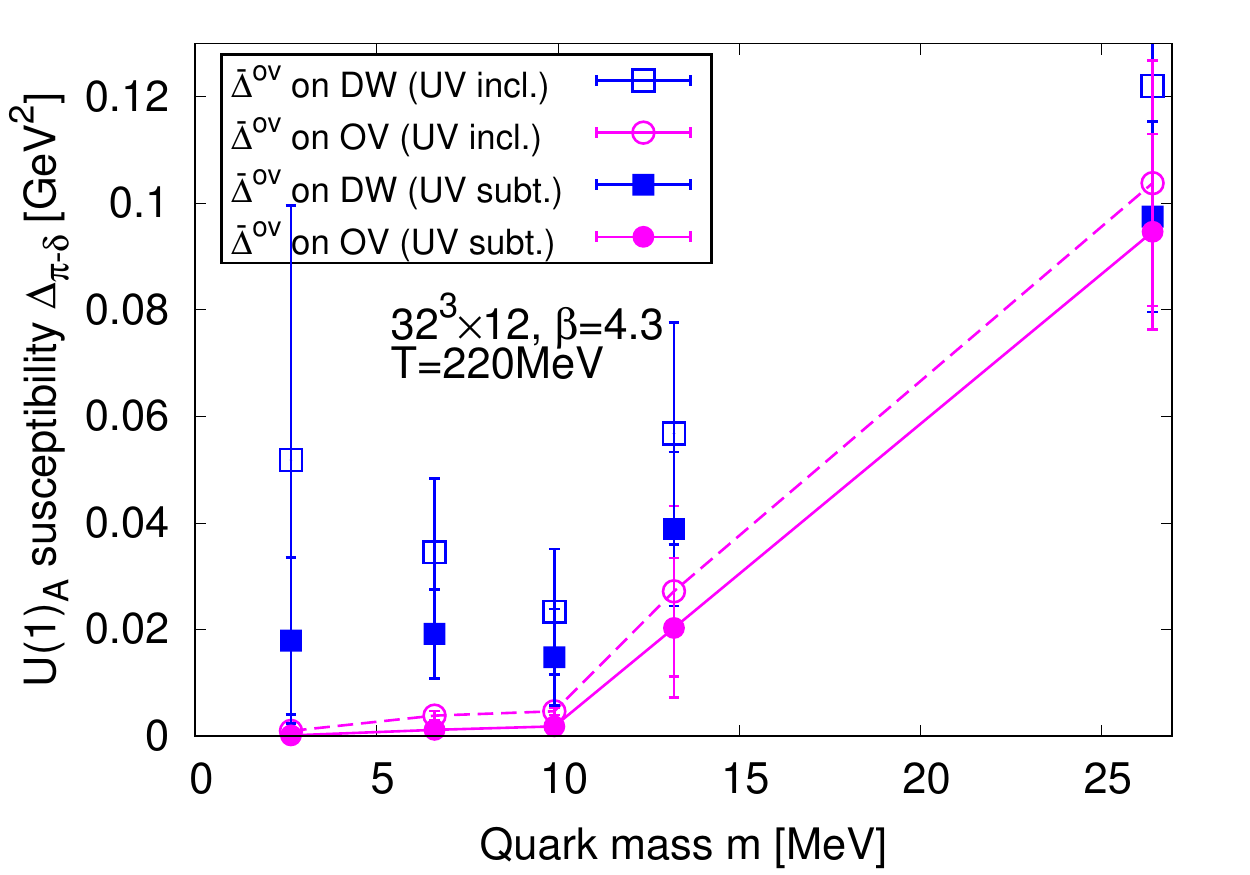}
    \end{center}
    \end{minipage}%
    \begin{minipage}[t]{0.5\columnwidth}
    \begin{center}
            \includegraphics[clip,width=1.0\columnwidth]{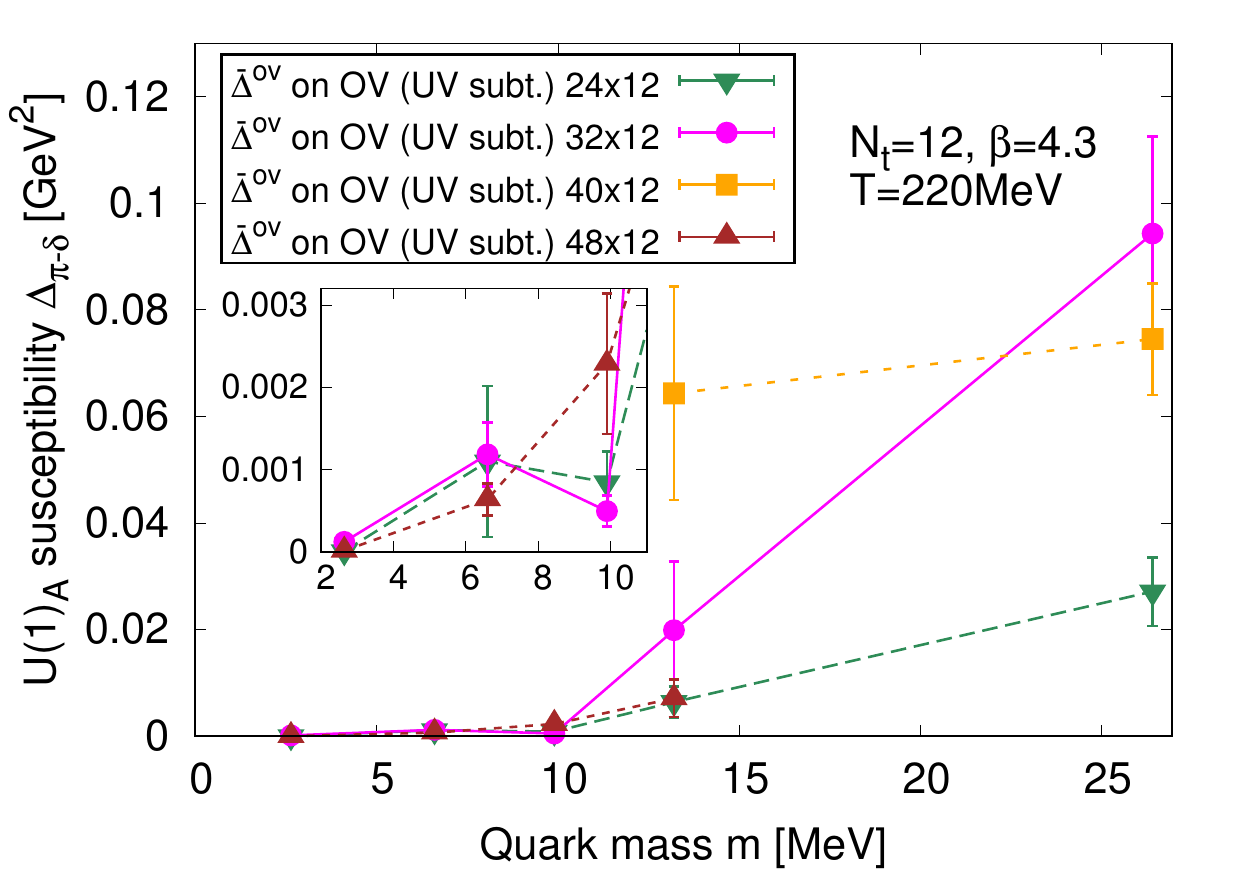}
    \end{center}
    \end{minipage}
    \caption{Quark mass dependence of $U(1)_A$ susceptibilities, $\bar{\Delta}_{\pi-\delta}^\mathrm{ov}$, from the eigenvalue density of the overlap-Dirac operators at $T=220 \, \mathrm{MeV}$.
Left: UV-included (open) or UV-subtracted (filled) results on the M\"obius domain-wall (squares) or reweighted overlap (circles) ensembles at $L=32$.
Right: Volume dependence ($L=24,32,40,48$).}
    \label{fig-2}
\end{figure}

In Fig.~\ref{fig-2}, we show the quark mass dependence of the $U(1)_A$ susceptibility $\bar{\Delta}_{\pi-\delta}^{\mathrm{ov}}$ at $T=220 \, \mathrm{MeV}$.
The left panel shows the results at the spatial volume $L^3=32^3$.
Here, the magenta circles (blue squares) represent the result on the OV (DW) ensembles.
$\bar{\Delta}_{\pi-\delta}^{\mathrm{ov}}$ on the DW suffers from fictitious modes by the violation of the GW relation, so that we expect that the results overestimate the true value.
On the other hand, $\bar{\Delta}_{\pi-\delta}^{\mathrm{ov}}$ on the OV is expected to be closer to the continuum limit.
Also, the open (filled) symbols denote the results before (after) the UV subtraction by the procedure of Eq.~(\ref{eq:Delta_subt}).
While the results with ultraviolet contributions overestimate $\bar{\Delta}_{\pi-\delta}^{\mathrm{ov}}$, the UV-subtracted results should be more reliable.
Therefore, in the following we focus on the filled magenta circles.

In the small quark mass region ($m \lesssim 10 \, \mathrm{MeV}$), $\bar{\Delta}_{\pi-\delta}^{\mathrm{ov}}$ nearly vanishes, which strongly suggests that the $U(1)_A$ symmetry is restored in the chiral limit.
Furthermore, near $m\sim 10 \, \mathrm{MeV}$, we find a sudden increase of $\bar{\Delta}_{\pi-\delta}^{\mathrm{ov}}$.
This behavior may imply the existence of a ``critical mass'' as discussed in Ref.~\cite{Aoki:2012yj}.
In the large quark mass region, $\bar{\Delta}_{\pi-\delta}^{\mathrm{ov}}$ shows a large value, which indicates that the $U(1)_A$ symmetry is clearly broken.

In the right panel of Fig.~\ref{fig-2}, we show the volume dependence of the $U(1)_A$ susceptibility.
For the small quark mass, there is no visible volume dependence between $L=24$ and $48$.
On the other hand, at the largest quark mass $m= 26.4 \, \mathrm{MeV}$, we found a clear volume dependence between $L=24$ and $32$.
We emphasize that the reason for this behavior needs to be carefully studied.



\subsection{Topological susceptibility}\label{subsec-3-3}

\begin{figure}[t!]
    \begin{minipage}[t]{0.5\columnwidth}
    \begin{center}
            \includegraphics[clip,width=1.0\columnwidth]{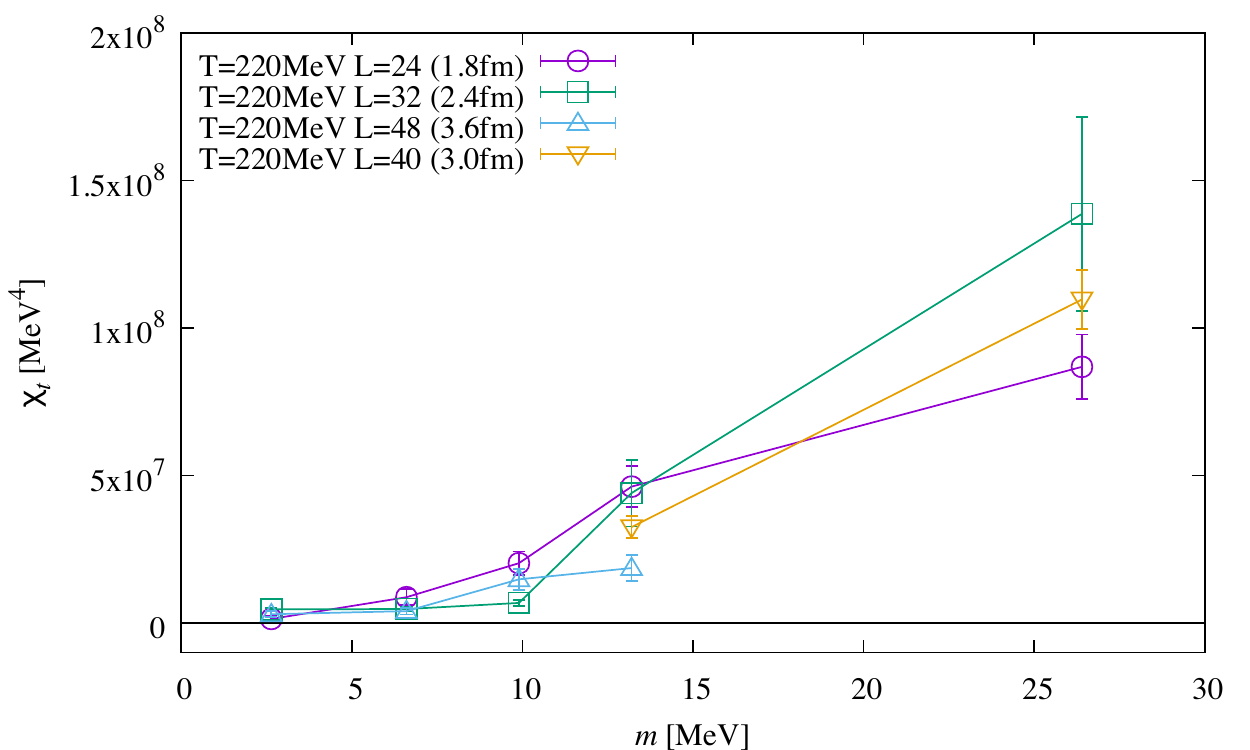}
    \end{center}
    \end{minipage}%
    \begin{minipage}[t]{0.5\columnwidth}
    \begin{center}
            \includegraphics[clip,width=1.0\columnwidth]{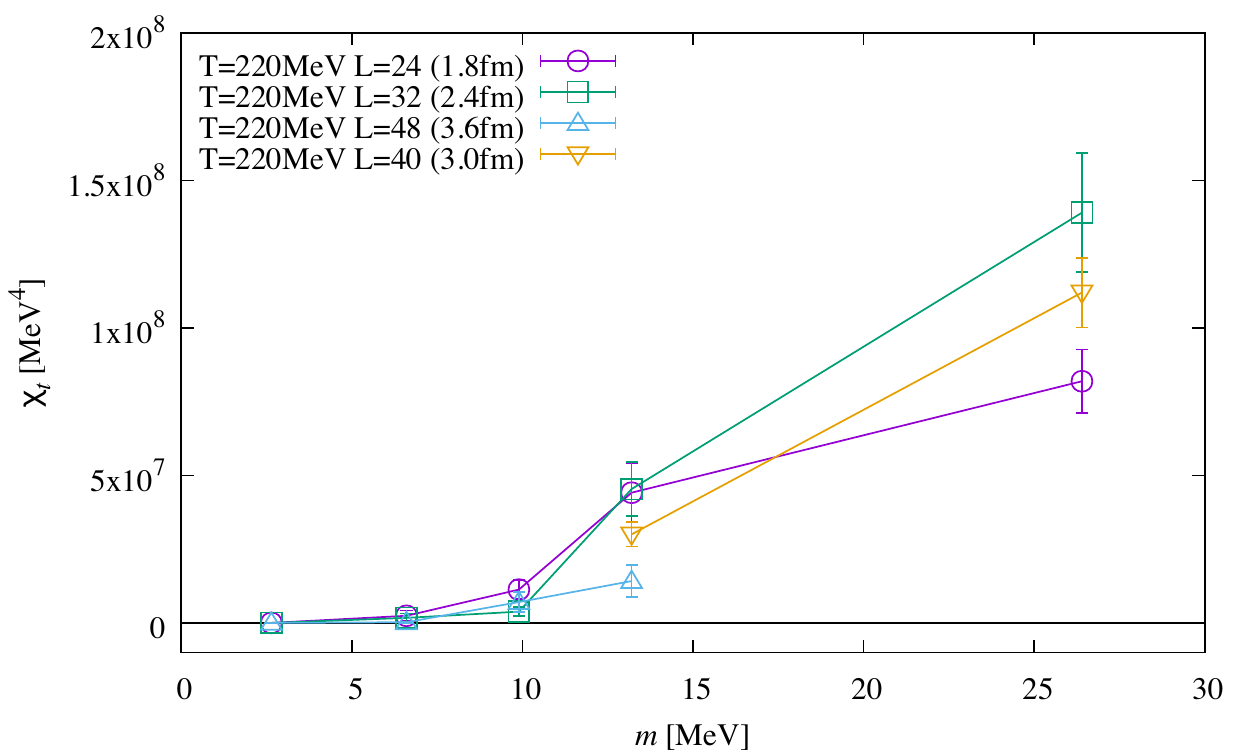}
    \end{center}
    \end{minipage}
    \caption{Quark mass and volume ($L=24,32,40,48$) dependence of topological susceptibilities $\chi_t$ at $T=220 \, \mathrm{MeV}$.
    Left: $\chi_t$ measured from gluonic operator on the M\"obius domain-wall ensembles.
    Right: $\chi_t$ measured from index of overlap Dirac operator on reweighted overlap ensembles.
}
    \label{fig-3}
\end{figure}

In Fig.~\ref{fig-3}, we show the quark mass dependence of the topological susceptibility $\chi_t$ at $T=220 \, \mathrm{MeV}$, where we compare the two types of measurements from the fermionic definition (\ref{Q_fermion}) or gluonic one (\ref{Q_gluon}) on the MDW or OV ensembles.
In the left panel, we focus on $\chi_t$ from the gluonic operator on the M\"obius domain-wall ensembles, and the right panel is from the index of overlap Dirac operator on reweighted overlap ensembles.
In the small quark mass region below $m\sim 10 \, \mathrm{MeV}$, both plots indicate that $\chi_t$ is strongly suppressed, and there is no visible volume dependence between $L=24$ and $48$ within the error bars.
On the other hand, in the large quark mass region, we find a nonzero value of $\chi_t$.

\section{Conclusion and outlook}\label{sec-4}
In this study, we investigated the $U(1)_A$ symmetry and topological charge in the high-temperature phase with $N_f=2$ lattice QCD simulation.
The quark mass dependence of the $U(1)_A$ susceptibility at $T=220 \, \mathrm{MeV}$ suggests the restoration of $U(1)_A$ symmetry in the chiral limit, which is consistent with the theoretical prediction of Ref.~\cite{Aoki:2012yj}.
As another observables for indicating the restoration of $U(1)_A$ symmetry, hadronic correlators are also interesting.
For example, the spatial mesonic correlators from our configurations are studied in Refs.~\cite{Rohrhofer:2017grg,Rohrhofer:2019qwq}.

In the future, simulations at lower temperature ($T<220 \, \mathrm{MeV}$) should be studied. 
In addition, the simulations including $N_f=2+1$ fermions need to be performed.
The previous studies with $N_f=2+1$ fermions in Refs.~\cite{Bazavov:2012qja,Buchoff:2013nra,Bhattacharya:2014ara,Dick:2015twa,Mazur:2018pjw} implied visible breaking of $U(1)_A$ symmetry, and the comparison between the results with $N_f=2$ and $N_f=3$ should be also important. 

\section*{Acknowledgment}\label{sec-5}
Numerical simulations are performed on IBM System Blue Gene Solution at KEK under a support of its Large Scale Simulation Program (No. 16/17-14) and Oakforest-PACS at JCAHPC under a support of the HPCI System Research Projects (Project ID:hp170061).
This work is supported in part by the Japanese Grant-in-Aid for Scientific Research (No. JP26247043, JP18H01216 and JP18H04484), and by MEXT as ``Priority Issue on Post-K computer" (Elucidation of the Fundamental Laws and Evolution of the Universe) and by Joint Institute for Computational Fundamental Science (JICFuS).
\bibliographystyle{JHEP}
\bibliography{lattice2018}
\end{document}